\documentclass{article}
\usepackage{subfigure,epsfig}
\usepackage{authblk}
\begin{document}

\title{Network Analysis of Urban Traffic 
\\with Big Bus Data}

\author{Kai Zhao}
\affil{New York University}
\date{}

\maketitle
\begin{abstract}

Urban traffic analysis is crucial for traffic forecasting systems, urban planning and, more recently, various mobile and 
network applications. 
In this paper, we analyse urban traffic with network and statistical methods. Our analysis is based on one big bus dataset containing 45 
million bus arrival samples in Helsinki. We mainly address following questions: 1. How can we identify the areas that cause 
most of the traffic in the city? 2. Why there is a urban traffic? Is bus traffic 
a key cause of the urban traffic? 3. How can we improve the urban traffic 
systems? To answer these questions, first, the betweenness is used to 
identify the most import areas that cause most traffics. Second, we find 
that bus traffic is not an important cause of urban traffic using statistical methods. 
We differentieate the urban traffic and the bus traffic in a city. We use bus delay as an identification of the urban traffic, and the number of bus as an identification of the bus traffic. Third, we give 
our solutions on how to improve urban traffic by the traffic simulation on 
road networks. 
We show that adding more buses during the peak time and providing better bus schedule plan in the hot areas like railway station, metro station, shopping malls etc. will reduce the urban traffic. 
\footnote{The technique report won the best hack award in Big Data Science Hackathon, Helsinki, 2015}

\end{abstract}




\section{Introduction}

Understanding urban traffic is crucial for traffic forecasting systems 
\cite{trafficone, traffictwo}, urban planning \cite{zhengyuubicomp2011, zhengyukdd2012} and, more recently, various mobile and 
network applications \cite{samulisensys13, transportationAppOne, 
transportationAppTwo, hui2009empirical, Mobisys15, Sensys13, SECON12}. 
We mainly address the following problems in this paper: 
\begin{itemize}
 \item RQ1. What area caused most of the traffic in the city? (network analysis methods) 
 \item RQ2. Why there is a urban traffic? Is bus traffic a key cause of the urban traffic? (statistical methods, correlation between bus traffic and 
 urban traffic)
 \item RQ3. How can we improve the urban traffic systems? (Simulation work)
 \end{itemize}
 
To answer these questions, first, we use the betweenness to 
identify the most import areas that cause most traffics. Second, we find 
that bus traffic is not an important cause of urban traffic using statistical methods. Third, we give 
our solutions on how to improve urban traffic by the traffic simulation on 
road networks. Adding more buses during the peak time and providing better bus schedule plan in the hot areas like railway station, metro station, shopping malls etc. will improving the Urban Transportation traffic.

First, we visualized the city as a network with bus stop as a node and the route between two bus stops as an edge. The edge weight is calculated by the average bus delay over two bus stops per hour. At first we calculated the average delay between two stops per hour and the number of buses passing through those stops in that hour. We then used this data to calculate the betweenness centrality. 

Second, we chose betweenness centrality to quantify the importance of the bus stops (nodes) in a network (City) by measuring the ratio of shortest paths passing through a particular node to the total number of shortest paths between all pairs of nodes. Betweenness centrality serves best in our quest to find most important stops in the road network since our purpose too is to identify the areas in the road network, which if jammed would have highest impact on overall city traffic. 

Third, we find that bus traffic is not an important cause of urban traffic using statistical methods. 
We find that the urban traffic is log-normal distributed and the bus traffic is power-law distributed. There is no correlation between the urban traffic and bus traffic using the Pearson correlation efficiency. Then, we give our solution on how to improve the urban traffic using simulations on the road networks.

\section{Overview}

\subsection{Dataset}

The publicly available HSL data is collected based on the run time of the services provided by the HSL in the Helsinki Area. It contains details about the Service Route, Service Vehicle Number, Expected Arrival Time, Expected Departure Time, Actual Arrival Time, Actual Departure Time etc. To understand the data better and get meaningful insights about the variables, we have done pre-processing and Exploratory Data Analysis. We have used histograms to know more about the data spread, boxplots to identify the outliers, Q-Q plots to identify the quantile ranges. The erroneous records in the data provided were discarded before further processing is done. The Bus delay emerged as the important covariate which explains more about the variability in the urban traffic. The Bus delay has been computed as the difference between the actual arriving time and arrival time according to the timetable between two bus stops. 

\subsection{Traffic delay as an identification of urban traffic}
\begin{figure}
\centering
\subfigure[Urban Traffic (Mon 8-9 am)]{
\includegraphics[width=0.30\textheight]{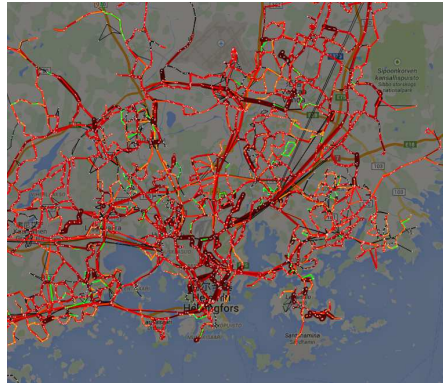}
}
\subfigure[Bus Traffic (Mon 8-9 am)]{
\includegraphics[width=0.30\textheight]{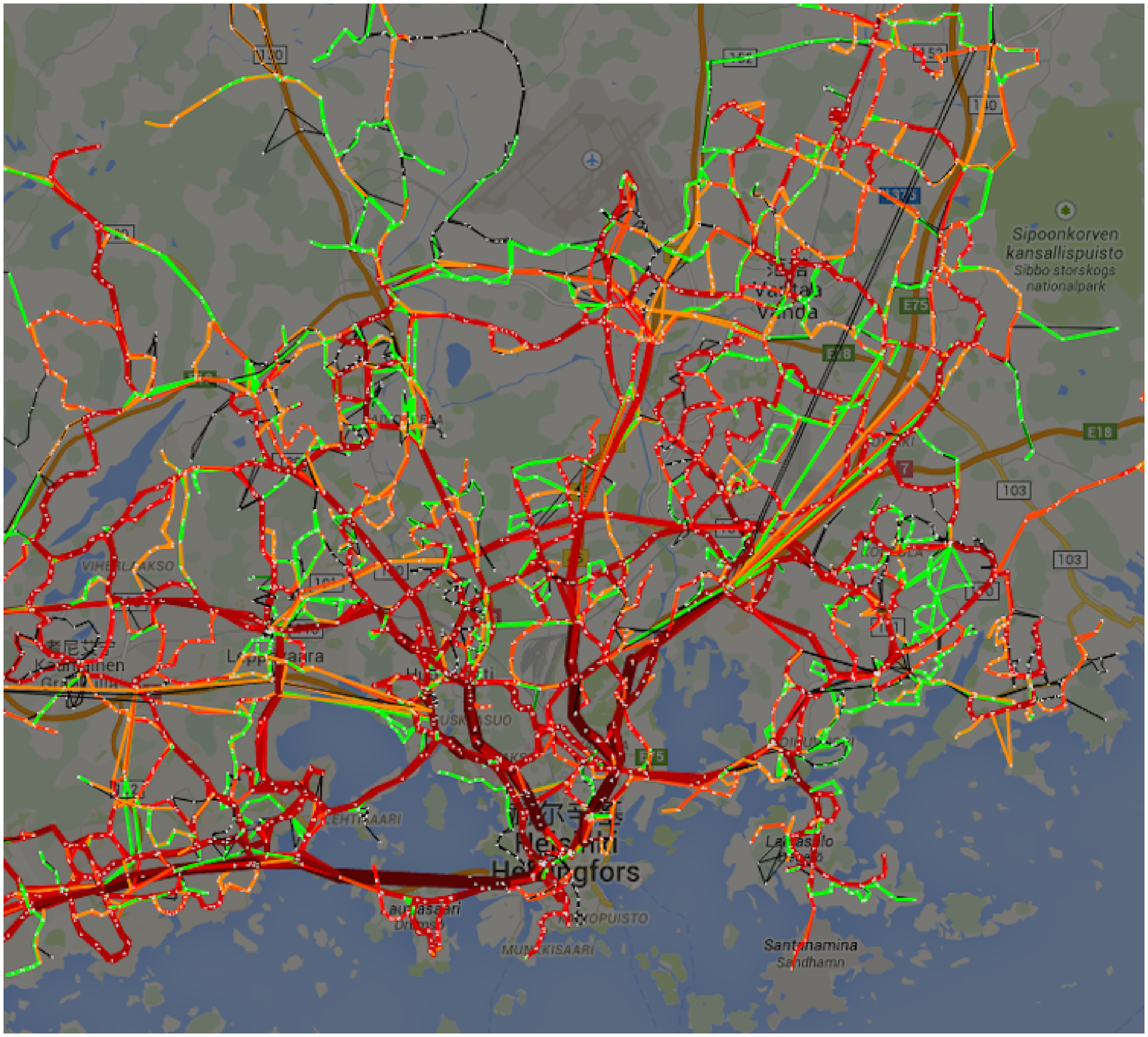}
}
\subfigure[Hot areas caused most traffic]{
\includegraphics[width=0.30\textheight]{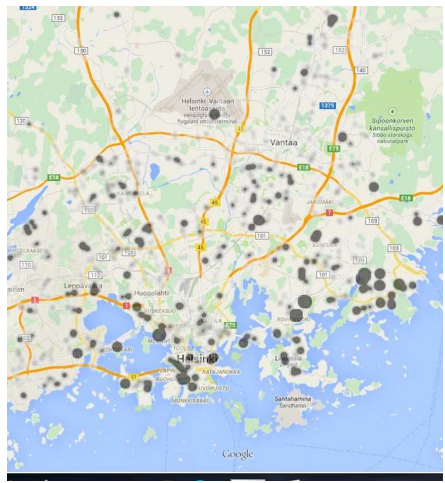} 
}

\caption{Network Analysis of Urban Traffic and Bus Traffic}\label{fig:traffic}
\end{figure}

Traffic delay is important aspect of our analysis, One can calculate current average delay between the stops by the HSL data. This if visualized, can allow one to figure out the busiest as well as fastest routes in the runtime. Since delay between two immediate stops is also available, a proper visualization can find the busiest/fastest  stretch within a route as well. For this, we consider the bus routes as a huge network with stops as their nodes and stretch between stops as their edges. The edges of the network are colored as a green-red gradient, where darker red means higher delay and green means close to zero delay. Once again we use Google maps service to draw up the road network of Helsinki, then after identifying the stops we color the edges between them with respective color. Fig. \ref{fig:traffic} a shows the result of this visualization. One can easily see the business of traffic in downtown as well as the  relatively low traffic in the outer part of the Helsinki. Instant availability of this data means visualization can be updated in the real time making traffic monitoring very easy. It could also provide information about sudden disruption in the traffic, for example if a relatively green stretch suddenly goes red this might be an indication of some sort of event at that place; may be an accident or road blockage.

\subsection{Urban Traffic Visualization}

We used Google maps service as the foundation of our visualization. It shows the basic map of the city along with the roads, we can identify the stops in the map with the help of longitude and latitude given in the HSL data. Thus we calculate the centrality of each stop; and then overlay a heatmap layer on the city map with a grey gradient. This way most important stops can be identified straight away by looking at the darkest blobs in the map. Fig. \ref{fig:traffic} a and b shows the resulting visualization of urban traffic and bus traffic over Helsinki city. One can immediately observe that the railway station, bigger intersections etc. have the darkest grey color blobs and hence have the utmost capability to disrupt overall traffic if they go down. 

\section{Hot areas that cause most of the traffic}

In this section, we use the betweenness to identify the hot areas that 
cause most of the traffic. Interesting part of a road network data set is that one can interpret it as a network with stops as its nodes and roads as its edges connecting stops to each other. This enables one to apply network analysis strategies to find out interesting properties of the network. We focused on two parts of the network; finding out the most important node and figuring out the busy/fast routes in the network. We provided a solution to improve the urban transportation by analysing the bus delays between two stops, the number of buses between those two stops and the correlation between bus traffic and urban traffic. Our idea is to analyse the urban traffic delay with the network traffic analysis methods.

\subsection{Identify hot areas}

We use centrality to identify the most important areas that cause the most 
traffic \cite{TMC15, ICDMW15, Thesis15}. Centrality indicator is the most common measure to find the important vertices in a network. Graph theory tells us that centrality can be calculated based on various metrics, i.e closeness, betweenness etc. For our analysis we use betweenness as the metric to find out the most important node. Betweenness centrality 
of a node is defined as the number of shortest paths from all vertices to all others passing through that node, thus a node with high betweenness will be the largest contributor to the efficient traffic management; it is also the  most sensitive point in the urban  traffic i.e. if it goes down, it causes huge disruption in the network. 

The betweenness centrality is defined as below:

\begin{equation}
C_B(v) = \sum_{s\neq_{v}\neq_{t}\in_{V}}{\frac{\sigma_{st}(v)}{\sigma_{st}}}
\end{equation}
 
The output obtained after applying the betweenness centrality is used to identify hot spots on the map. Once the centrality of all the nodes is calculated, one need to display it in a way so that not only the information can be grasped easily but also the context of the analysis is not lost in technical details. Our analysis was to discover the nodes with the highest centrality with respect to the betweenness,i.e figuring out the most important bus stops, which if jammed would disrupt the traffic on most of the routes (Figure \ref{fig:traffic} c). Thus it makes sense to overlay this information over the geographic map of the city, showing up the road network. Another aspect was to identify the pair of stops with highest/lowest delay; these would be the edges with in the road network. This information would again overlay on the city road network.

\section{Correlation between Bus Traffic and Urban traffic} 

In this section we mainly use statistical methods to analyse the correlation between urban traffic and bus traffic. We find that 
bus traffic is not a key cause of urban traffic. To study the cause of bus delay and urban traffic we studied correlation between bus delay and the number of buses. The bus delay between two stops being urban traffic and number of buses travelling through that two stops being bus traffic. We choose two peak times for this purpose which are Monday 8 AM- 9 AM and Monday 4 PM – 5 PM as shown in Figure \ref{fig:sta}.

\subsubsection{Fitting distributions}

To find the model that fits our data \cite{SR15}, we used Akaike's information criterion (AIC), in combination with Maximum likelihood estimation (MLE). AIC is used to identify the best fitting distribution among all fitted distributions and MLE is used to find an estimator that maximizes the likelihood function of one distribution.
\begin{equation}
AIC = -2log\left( L\left( \hat{\theta}|data\right) \right) +2K 
\end{equation}

The AIC value of each fitted distributions are normalised by calculating the delta AIC between different AIC values which is a measure of each distribution relative to the best distribution, and is calculated as
\begin{equation}
\bigtriangleup_i = AIC_i - AIC_{min}
\end{equation}
Akaike weights are then calculated to measure of the strength of evidence for each distribution and is given as, 
\begin{equation}
W_{i} = \frac{exp\left( -\bigtriangleup_{i}/2\right) }{\sum \limits_{r=1}^{R}exp\left({-}\bigtriangleup_{i}/2\right)  }
\end{equation}

\begin{figure*}
\centering

\subfigure[Urban Traffic (Mon 8-9 am) follow log-normal distribution]{
\includegraphics[width=0.27\textheight]{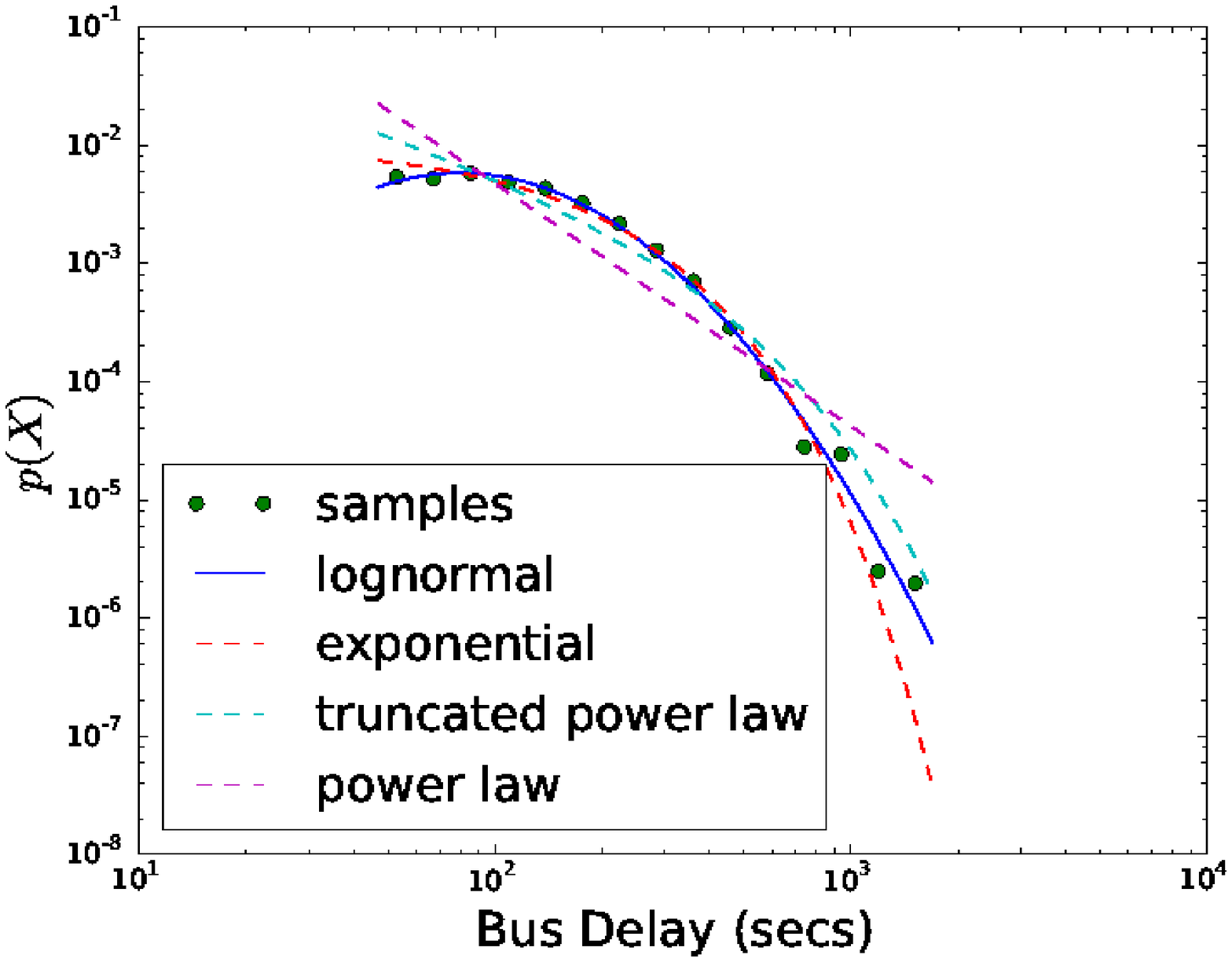}
}
\subfigure[Bus Traffic (Mon 8-9 am) follow power-law distribution]{
\includegraphics[width=0.27\textheight]{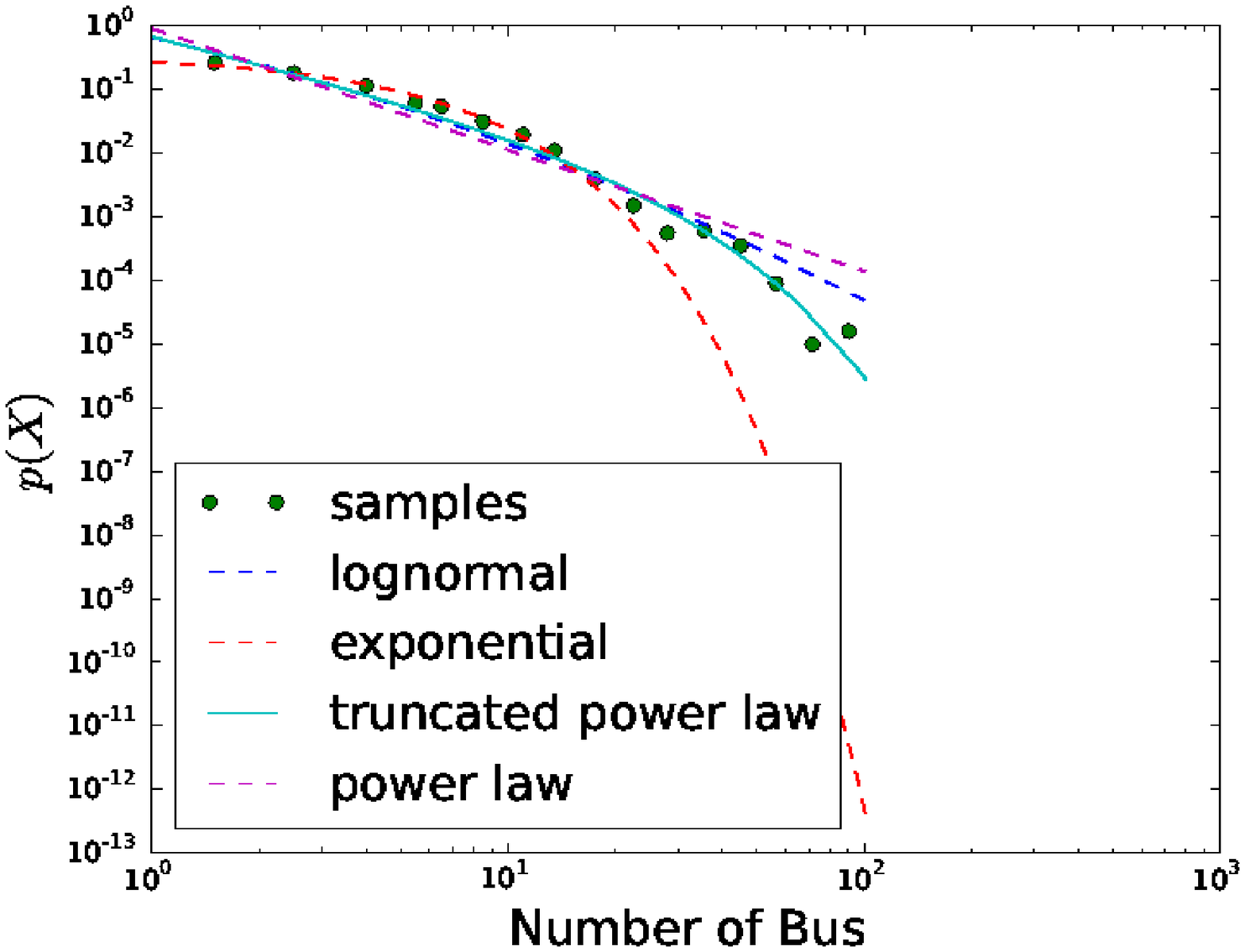}
}
\subfigure[Urban Traffic (Mon 4-5 pm) follow log-normal distribution]{
\includegraphics[width=0.27\textheight]{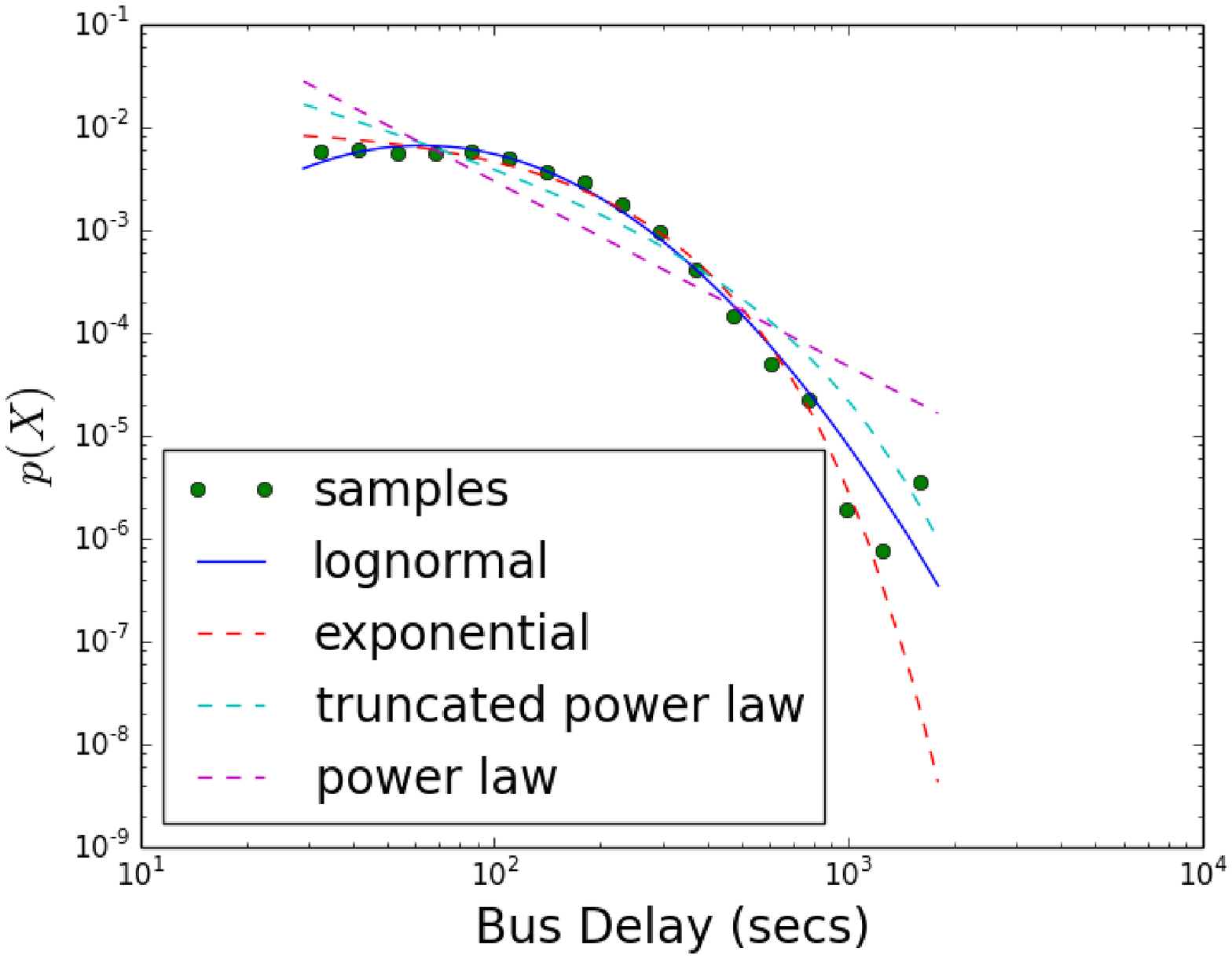}
}
\subfigure[Bus Traffic (Mon 4-5 pm) follow power-law distribution]{
\includegraphics[width=0.27\textheight]{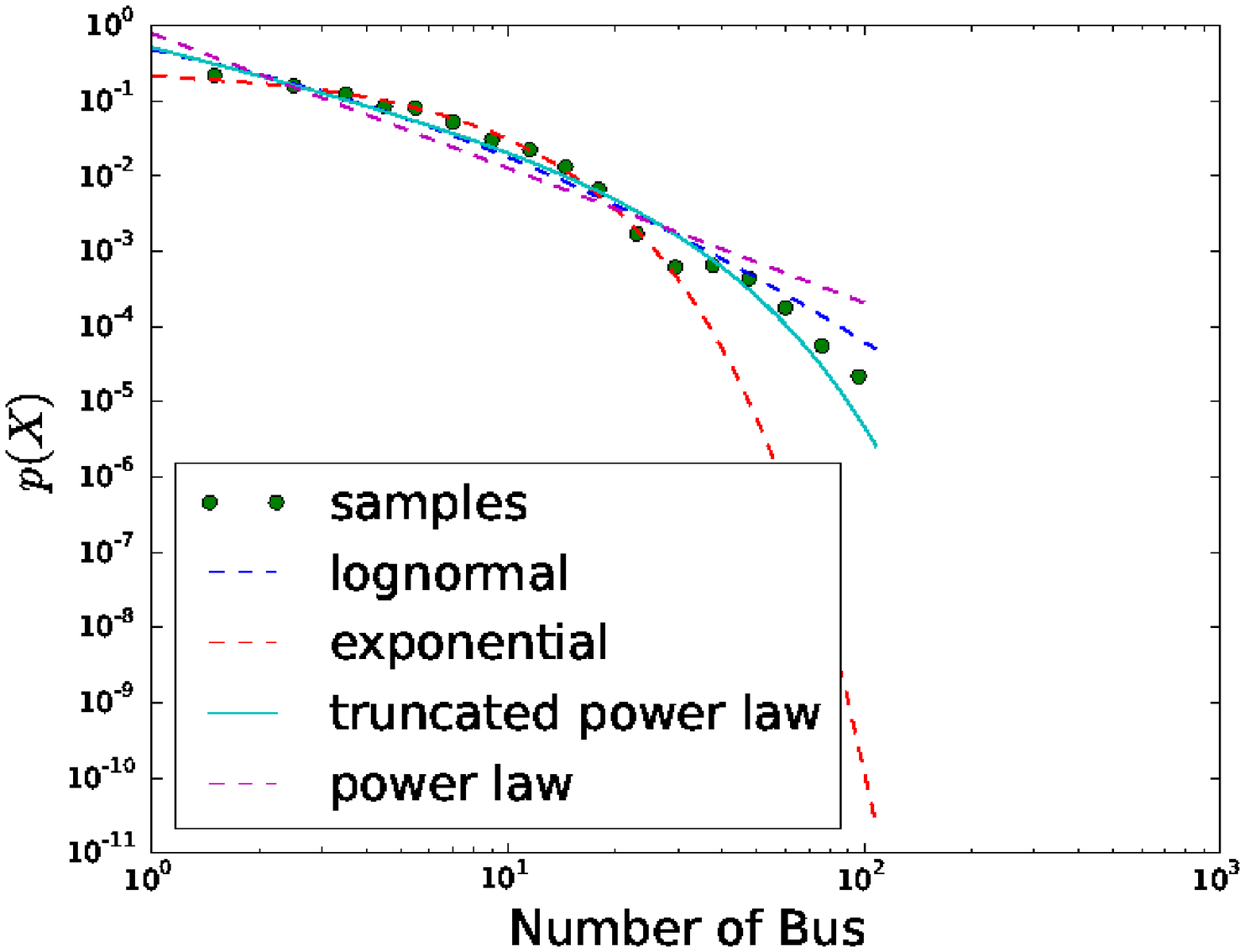}
}
\caption{Distribution of urban traffic and bus traffic.}\label{fig:sta}
\end{figure*}

We used the following distributions for the study and their corresponding Probability Density Function is mentioned below,
Truncated Pareto distribution with probability density function of
\begin{equation}
Cx^{-\alpha}e^{-\lambda{x}}
\end{equation}  
Log-normal distribution with probability density function of 
\begin{equation}
\frac{1}{x\sigma\sqrt{2\pi}}exp\left[- \frac{\left( \ln\left( x\right) - \mu\right) ^2}{2\sigma^2}\right] 
\end{equation}
Pareto distribution with probability density function of 
\begin{equation}
\left( \alpha-1\right) x_{min}^{\alpha-1}x^{-\alpha}
\end{equation}
Exponential distribution with probability density function of 
\begin{equation}
\lambda e^{-\lambda{x}}
\end{equation}
From the study we found that the Lognormal distribution fits the Urban Traffic, and power-law fits the bus traffic. The Fig. \ref{fig:sta} 
a. and c. corresponds to the lognormal distribution of the urban traffic and Fig. \ref{fig:sta} b. and d. corresponds to the power-law disributions of the bus traffic.

To measure the strength of the correlation between bus traffic and urban traffic we used Pearson correlation which is given by:
\begin{equation}
\rho_{X,Y} = \frac{E\left[ \left( X - \mu_X \right)\left( Y - \mu_Y\right)  \right] }{\sigma_X \sigma_Y}
\end{equation}

\begin{figure*}
\centering

\subfigure[Correlation between Urban Traffic and Bus Traffic (Mon 8-9 am)]{
\includegraphics[width=0.27\textheight]{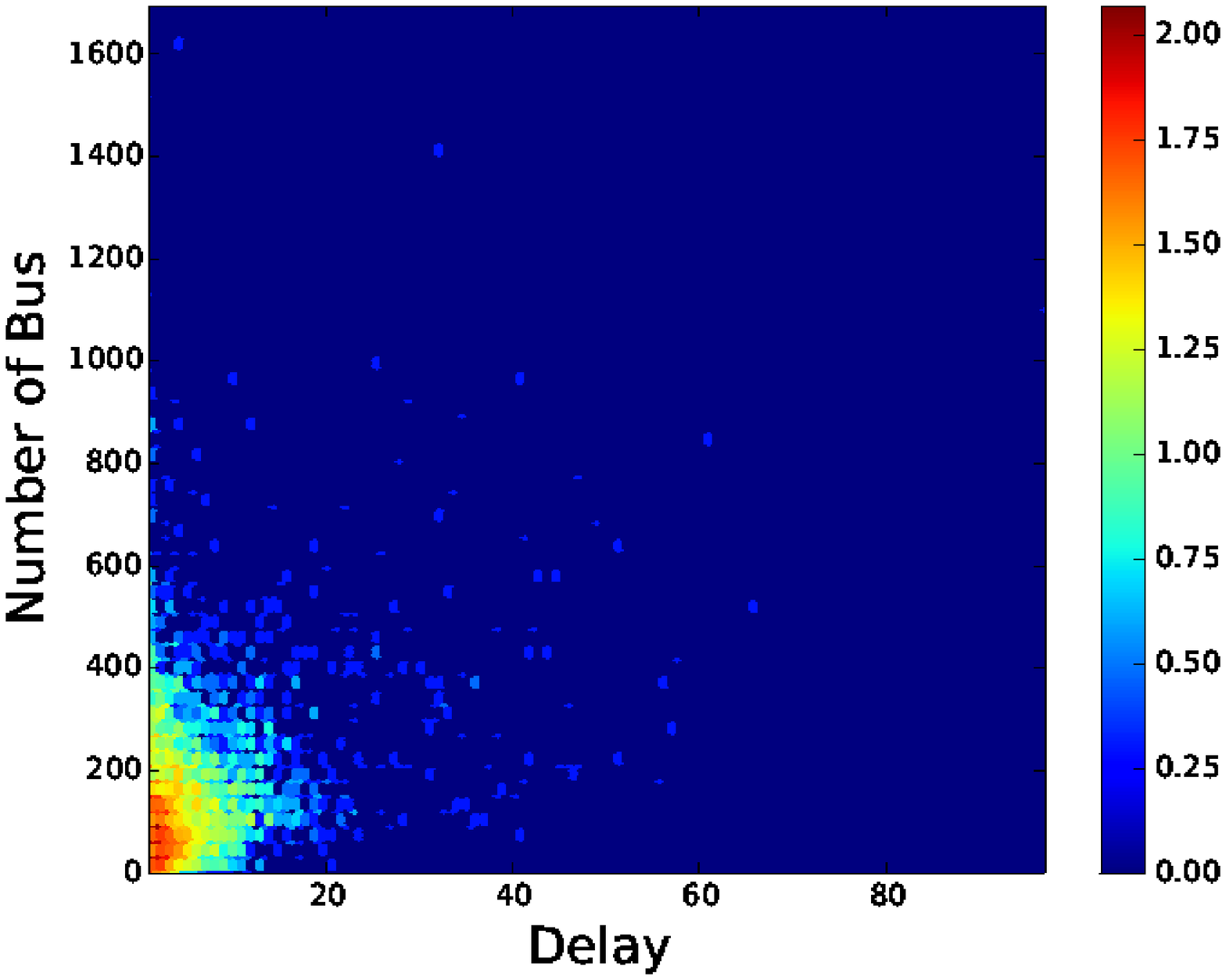} 
}
\subfigure[Correlation between Urban Traffic and Bus Traffic (Mon 4-5 pm)]{
\includegraphics[width=0.27\textheight]{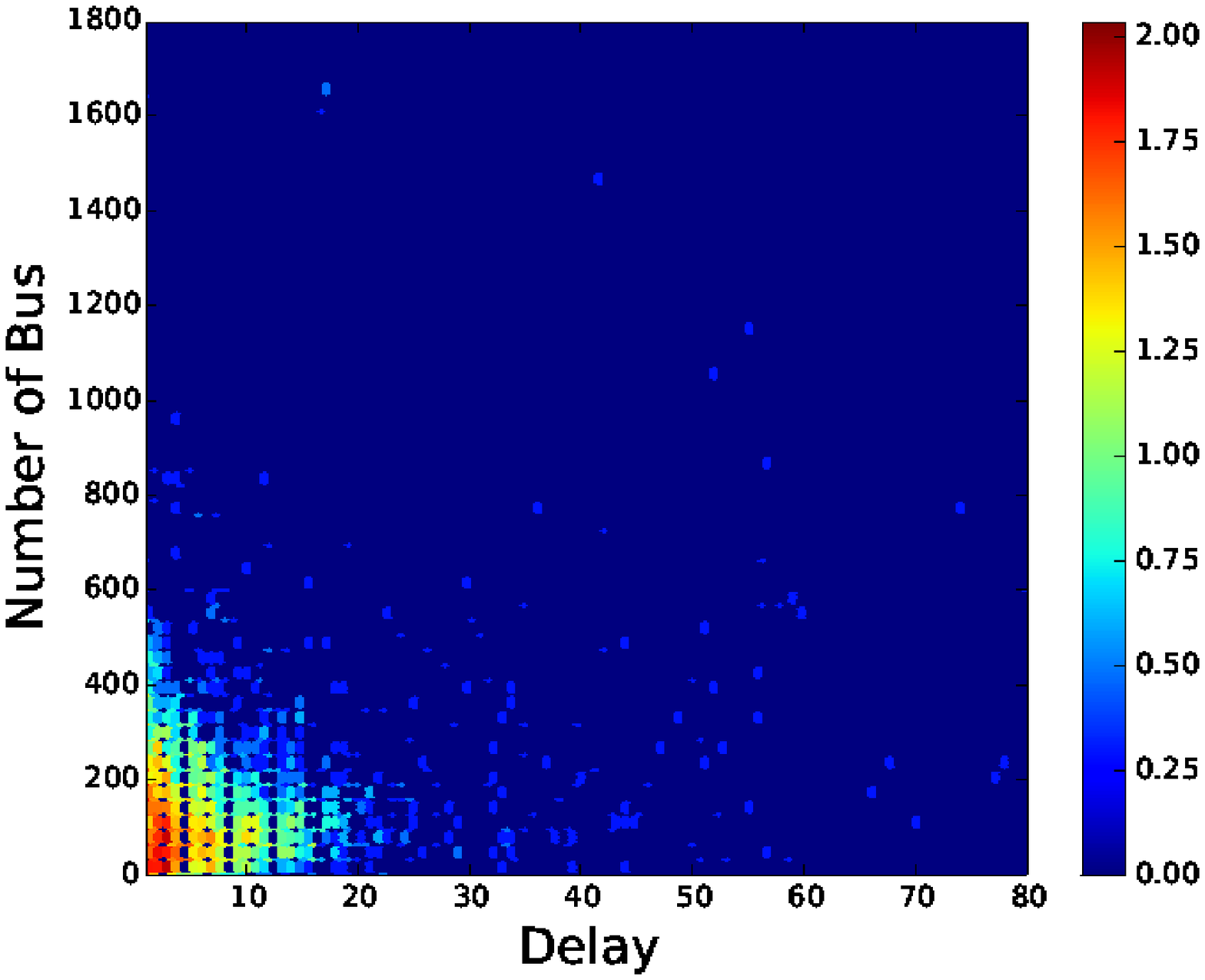} 
}
\caption{Correlation between Urban Traffic and Bus Traffic}\label{fig:sta1}
\end{figure*}

The Pearson correlation plot as shown in the Fig. \ref{fig:sta1} shows, that there is no correlation between bus traffic and the urban traffic. 

We observed Pearson values of 0.22403 and 0.13301 in the morning and afternoon, respectively. The Pearson values that we observed clearly shows that there is no linear correlation between bus traffic and urban traffic.
From the study we observed that the urban traffic is not only due to bus traffic. That is the bus delay between two stops is not only due to the number of buses running between those two corresponding stops, and is also due to other factors like vehicular traffic, traffic signal, number of passengers getting on and getting off the bus, etc.

\section{How can we improve urban traffic systems}

In this section we show how we can increase the traffic throughput in 
the city environment by the simulation of human movement in the cities. 
Our simulation work shows that, for improving the Urban Transportation systems is to add more buses during the peak time, 
reduce other vehicle usage, reduce pick-up and drop time, and finally provide better bus schedule plan in the hot areas like railway station, metro station, shopping malls etc. 

\section{Conclusion}
Our analysis of open bus data shows that in city urban traffic is not caused by the bus traffic alone, i.e the delay in the stretches is also due to the private traffic. Thus, increasing buses on routes would not increase the delay in the routes. Additionally we presented a way to find out and visualize most important bus stops in the region operates, this information combined with the real time delay visualization can help in figuring out the pain points of Helsinki traffic network in real time, for example An emerging delay in an area of high centrality must be addressed quickly as it has potential to spill over to the other parts of network. Thus methodology used in this analysis can enable people to figure out the potential disruptions in the traffic in real time, classify them as usual/unusual and address them quickly on a day to day basis. Our solution to improve the Urban Transportation systems is to add more buses during the peak time, reduce other vehicle usage, reduce pick-up and drop time, and finally provide better bus schedule plan in the hot areas like railway station, metro station, shopping malls etc.

\bibliographystyle{abbrv}
\bibliography{thesis} 

\end{document}